\documentclass[10pt, conference, compsocconf]{IEEEtran}
\IEEEoverridecommandlockouts
\usepackage{cite}
\usepackage{amsmath,amssymb,amsfonts}
     \usepackage{amsthm}    
\usepackage{algorithmic}
\usepackage{graphicx}
\usepackage{textcomp}
\def\BibTeX{{\rm B\kern-.05em{\sc i\kern-.025em b}\kern-.08em
    T\kern-.1667em\lower.7ex\hbox{E}\kern-.125emX}}

\usepackage[utf8]{inputenc}
\usepackage{tikz}
\usepackage[english]{babel}
\usepackage{dirtytalk}
\usepackage{amsmath}
\usepackage{amssymb}
\usepackage{pifont}
\usepackage{mathtools}
\usepackage{float}
\usepackage{graphics}
\usepackage{color}
\usepackage{cancel}
\usepackage{graphicx} 
\usepackage{caption}
\usepackage[ruled,vlined]{algorithm2e}
\usepackage{tkz-graph}
\usepackage{tabularx}
\usepackage{multirow}
\usepackage{multicol}
\usepackage{csvsimple,booktabs,siunitx}
\usepackage{adjustbox}
\usepackage{xifthen}
\usepackage[normalem]{ulem}
 \usepackage{subfig}
 \usepackage{booktabs,makecell}

  \usepackage[hidelinks,bookmarks=false]{hyperref}
  
\usetikzlibrary{positioning}
\usetikzlibrary{matrix,backgrounds}
\usetikzlibrary{arrows,automata}
\usetikzlibrary{shapes,shapes.geometric,fit,calc}
\usetikzlibrary{decorations.pathreplacing}
\usetikzlibrary{patterns}
\GraphInit[vstyle = Normal]
\tikzset{
  VertexStyle/.append style = { inner sep=2pt,
                                font = \bfseries},
  EdgeStyle/.append style = {->} }

  \newtheorem{definition}{Definition}
    \newtheorem{example}{Example}

\newcommand{\ffdlt}[1][l]{%
	\ifthenelse{\equal{#1}{l}}{Friend-Foe Dynamic Linear Threshold Model\xspace}{$F^2DLT$}%
}

\newcommand{\ffdlts}{Friend-Foe Dynamic Linear Threshold Models}

\newcommand{\noncomp}[1][l]{%
	\ifthenelse{\equal{#1}{l}}{Non-Competitive Friend-Foe Dynamic Linear Threshold Model\xspace}{$nC\text{-}F^2DLT$}%
}

\newcommand{\semiprog}[1][l]{%
	\ifthenelse{\equal{#1}{l}}{Semi-Progressive Competitive Friend-Foe Dynamic Linear Threshold Model\xspace}{$spC\text{-}F^2DLT$}%
}

\newcommand{\nonprog}[1][l]{%
	\ifthenelse{\equal{#1}{l}}{Non-Progressive Competitive Friend-Foe Dynamic Linear Threshold Model\xspace}{$npC\text{-}F^2DLT$}%
}

\newcommand{\as}[2]{
	\ifthenelse{\equal{#1}{+}}{S'_{#2}}{}
	\ifthenelse{\equal{#1}{-}}{S''_{#2}}{}
	\ifthenelse{\isempty{#1}}{S_{#2}}{}
}

\newcommand{\aqs}[2]{
	\ifthenelse{\equal{#1}{+}}{\widetilde{S'}_{#2}}{}
	\ifthenelse{\equal{#1}{-}}{\widetilde{S''}_{#2}}{}
	\ifthenelse{\isempty{#1}}{\widetilde{S}_{#2}}{}
}

\def\G{\mathcal{G}}

\begin{document}
\title{Trust-based dynamic linear threshold models for non-competitive and competitive influence propagation}

\author{\IEEEauthorblockN{Antonio Cali\`{o}}
\IEEEauthorblockA{\textit{DIMES, University of Calabria} \\
87036 Rende (CS), Italy \\
 a.calio@dimes.unical.it}
\and
\IEEEauthorblockN{Andrea Tagarelli}
\IEEEauthorblockA{\textit{DIMES, University of Calabria} \\
87036 Rende (CS), Italy \\
 andrea.tagarelli@unical.it}
}

\maketitle 
\begin{abstract}
What are the key-features that enable an information diffusion model to explain the inherent dynamic,  and often  competitive, nature of real-world  propagation phenomena? In this paper we aim to answer this question by proposing a novel class of diffusion models, inspired by the classic Linear Threshold model, and built around the following aspects:  trust/distrust   in the user relationships, which is leveraged to model different effects of social influence on the decisions taken by an individual;  changes  in adopting one or alternative information items;   hesitation towards    adopting an information item over time;   latency in the propagation; time horizon for the unfolding of the diffusion process; and multiple cascades of information that might occur competitively. 
To the best of our knowledge, the above aspects have never been unified into the same LT-based diffusion model. 
 We also define different strategies for the selection of the initial influencers  to simulate non-competitive and competitive diffusion scenarios,   particularly related to the problem of limitation of misinformation spread. Results on publicly available  networks have shown the meaningfulness and uniqueness of our  models.  
\end{abstract}

\begin{IEEEkeywords}
information diffusion, influence propagation,  trust/distrust relationships, limitation of misinformation spread 
\end{IEEEkeywords}

\section{Introduction}

Since the early applications in viral marketing, 
the development of information diffusion models and their   embedding in optimization methods has provided effective support to address a variety of influence propagation problems. However, one criticism that arises from  existing diffusion models is the concern as to whether, and to what extent, they are adequate to explain the actual \textit{complexity of influence propagation  phenomena} that occur in the modern society of information.   
 The acquisition and share of true or reliable  information has  often to cope with unlimited \textit{misinformation} spots over the Web, e.g., \textit{fake news} ~\cite{Kumar:2016}  
mostly associated with 
 consequences on the real life of individuals.  
 %
 %
A few studies on the spreading of fake news and hoaxes (e.g.,~\cite{metaxas1})  
    have found that the difficulty for users of checking  the reliability or trustworthiness of the  web source generating and/or sharing the information item, can  increase the likelihood of people to be deceived. 
Within this view, one side effect is the tendency of users to access information from like-minded sources 
 and remain within information bubbles. 
 In general, in the attempt of debunking misinformation, one might intuitively recognize 
 two main strategies: real-time detection and correction, or delayed correction~\cite{Kumar2014DetectingMI}. In both cases, the response time plays a crucial role into the limitation of  misinformation diffusion, since 
 users tend  to reinforce their own belief --- a cognitive phenomenon known as \textit{confirmation bias}. 
  It may also happen that 
 such corrections do not yield the expected outcome, 
 or   they  may even produce \say{backfire} results 
 driving people’s attention to the fake news. 

In this scenario, it appears  that one recipe to deal with the interleaving of information and dis/misinformation should be to educate people to be mindful of the informative source. Unfortunately, 
 it is often difficult to understand where an information item originated from.
  Therefore, it turns out to be  essential to capture the effects that  different types of social ties, particularly \textit{trust/distrust relationships},   can have on both the user behavior and propagation dynamics. 
Two related questions hence arise:   
\textbf{Q1} -- {\em What are the key-features that make a diffusion model able to explain the inherent  dynamic, and often competitive, nature of real-world  propagation phenomena?} \   
\textbf{Q2} -- {\em Do  the currently used models of diffusion already incorporate such features?} 
 %

%

To address question \textbf{Q1}, we recognize   a number of  aspects   as essential constituents of a ``realistic'' information diffusion model, namely:  
 %
(1) leveraging trust/distrust information in the user relationships to  capture different effects of influence on decisions taken by a user;  
(2) accounting for  a user's change  in adopting one or alternative   information items (i.e.,  relaxation of the diffusion progressivity assumption);  
(3) accounting for a user's hesitation or inclination towards the  adoption of an information over time; 
(4) accounting for  time-dependent variables, such as latency, to explain  the propagation dynamics; 
(5) dealing with multiple cascades of information that might occur competitively.
%

\textbf{Motivating example.\ }    
To support our above hypothesis with an example, consider a typical scenario occurring in a political campaign, where two candidates want to target the audience of potential electors. Assume that every elector is initially unbiased toward one of the two candidates.  
 The decision about which candidate to vote it will likely be  the result of  exogenous and endogenous influencing factors, i.e., one  may be genuinely influenced by decisions taken by her/his social contacts --- impact of  homophily factors --- but s/he may also have formed her/his own opinion outside the network of friends.  
However, an individual's decision can also be influenced by the behavior of  neighboring \textit{foes}.  As a consequence of such negative influence received by foes, one may become more hesitant in taking a decision, which would be reflected by a \textit{quiescence} status of the elector before being fully engaged in the promotion of the chosen candidate.  
Moreover, once an  elector becomes active in favor of a particular candidate, it will be more difficult to change her/his mind over time, therefore a time-aware notion of \textit{activation threshold} is needed to model the effects due to the \textit{confirmation bias}. 
Finally, all decisions must be taken before the time limit (i.e., the election day) that constrains the political campaign period.

\begin{table}[t!]
\centering
\caption{Summary of  related work based on    optimization problem, basic diffusion model (\textbf{DM}), competitive diffusion (\textbf{C}), non-progressivity (\textbf{NP}),   time-aware activation (\textbf{TA}), delayed propagation (\textbf{DP}),   trust/distrust relations (\textbf{TD}).}
\label{tab:related}
\scalebox{0.77}{
\begin{tabular}{|l|c|c|c|c|c|c|c|}
\hline
\textbf{Ref.}	 & 	\textbf{Problem}	 & 	\textbf{DM}	 & 	\textbf{C} 	 & 	\textbf{NP}	 & 	\textbf{TA}	 & 	\textbf{DP} 	 & 	\textbf{TD}	\\
\hline \hline
  \cite{Budak+2011} & 	rumor blocking  & 	 IC	 & 	\checkmark	 & 	 	 & 	 	 & 	 	 & 	 	\\
 \cite{Anefficient}	 & rumor blocking 	 & 	IC 	 & 	\checkmark	 & 	 	 & 		 & 		 & 		\\
\cite{He2012463}	 & 	rumor blocking	 & 	LT	 & 	\checkmark	 & 	  & 		 & 		 & 		\\
\cite{Fan2013540}	 & 	rumor blocking	 & 	\footnotesize{distrib.} & 	\checkmark	 & 	 	 & 		 & 		 & 		\\
\cite{Chen2011379}	 & 	positive influ. max.	 & 	IC 	 & 	\checkmark	 & 	 	 & 		 & 	 	 & 	 	\\
\cite{Lou2014131} & active time max. & IC & & \checkmark & & & \\
\cite{Fazli2012315} & PTS min. & LT & & \checkmark & & & \\
%
%
\cite{Pagerank}	 & 	positive influ. max.  	 & 	Voter  	 & 	\checkmark	 & 		 & 		 & 	 	 & 	\checkmark	\\
\cite{Talluri2015479} & 	positive influ. max. & 	 LT	 & 	\checkmark	 & 		 & 		 & 	 	 & 	\checkmark	\\
\cite{Weng20171931} & 	positive influ. max.	 & 	 LT 	 & 	 \checkmark	 & 		 & 		 & 	 	 & 	\checkmark	\\
\cite{Liu2012439}	 & 	time-constrain.  influ. max.  & 	IC 	 & 	 	 & 	 	 & 		 & 	\checkmark	 & 	 	\\
\cite{Chen2012592}	 & 	time-constrain. influ. max. 	 & 	IC 	 & 	 	 & 	 	 & 		 & 	\checkmark & 	 	\\
\cite{MohamadiBaghmolaei2015195} & 	positive influ. max.	 & 	IC 	 & 	\checkmark	 & 	\checkmark & 		 & 	\checkmark 	 & 	\checkmark	\\
\cite{realtimecosteffective}	 & 	rumor blocking	 & 	LT	 & 	\checkmark	 & 	\checkmark 	 & 	\checkmark	 & 	 	 & 	 	\\
\cite{Lu:2015}	 & 	positive influ. max.	 & 	IC	 & 	\checkmark	 & 	 	 & 		 & 	 	 & 	 	\\
\hline
\end{tabular} 
 }
\end{table}
 
Concerning   question \textbf{Q2}, 
 a relatively large corpus of research studies has been developed in the last few years in the attempt of  explaining realistic  propagation phenomena, building upon     classic  information diffusion models, such as   Independent Cascade (IC) and Linear Threshold  (LT)~\cite{Kempe2003137}. 
   Table~\ref{tab:related} provides a schematic overview of models that 
incorporate one or more of the  aspects mentioned before about \textbf{Q1};  it is worth noting that  no existing work unifies \textit{all} of the above aspects into the same   (LT-based) diffusion model.

\textbf{Contributions.\ } 
In this paper, we propose a novel class of diffusion models, named \textit{\ffdlts} (\ffdlt[s]), 
which  are based on the classic LT model and are designed to deal with \textit{non-competitive} as well as \textit{competitive} time-varying propagation scenarios. 
In our proposed models,  
 the information diffusion graph is defined on top of a trust network, so that the strength of trust and distrust relationships is encoded into the influence probabilities. 
The behavior of a user in response to influencing actions  is modeled with a time-varying activation function, depending on both the inherent  activation-threshold of the user and her/his tendency of keeping or leaving the   campaign-specific  activation state over time. We also introduce a  quiescence function  to model the latency or delay that the influence of foes may determine in the   participation of a user in the information propagation.  
  For competitive scenarios, we define   a \textit{semi-progressive} model, which assumes that a user, once activated, is only allowed to switch to a different campaign, and a \textit{non-progressive} model, which instead requires a user to have always the support of her/his in-neighbors to keep the activation state with a certain campaign. 

Another contribution of this work is the definition of  four \textit{seed selection} strategies,     which mimic  different, realistic scenarios of influence propagation.  These strategies are central to our methodology of propagation simulation, since the development of optimization methods  under our diffusion models is beyond the goals of this work.  Notably, in competitive scenarios,  we have focused on combinations of  campaign strategies  that  might be reasonably considered for a misinformation spread limitation problem.   
Experimental evaluation conducted on four real-world networks has provided interesting findings on the meaningfulness and uniqueness    of our proposed models.

\section{\ffdlts} 
\label{sec:TTM}
Here we describe our proposed  \ffdlt[s]   models:   
the Non-Competitive \ffdlt[s] (\noncomp[s]), 
the  Semi-Progressive \ffdlt[s] (\semiprog[s]), and the Non-Progressive \ffdlt[s] (\nonprog[s]). 
 We first provide an overview of the   framework based on \ffdlt[s].  Next, we   introduce key features common to all   models, then we elaborate on each of them.

\subsection{Overview}
\begin{figure}[t!]
\centering 
\includegraphics[width=0.45\textwidth]{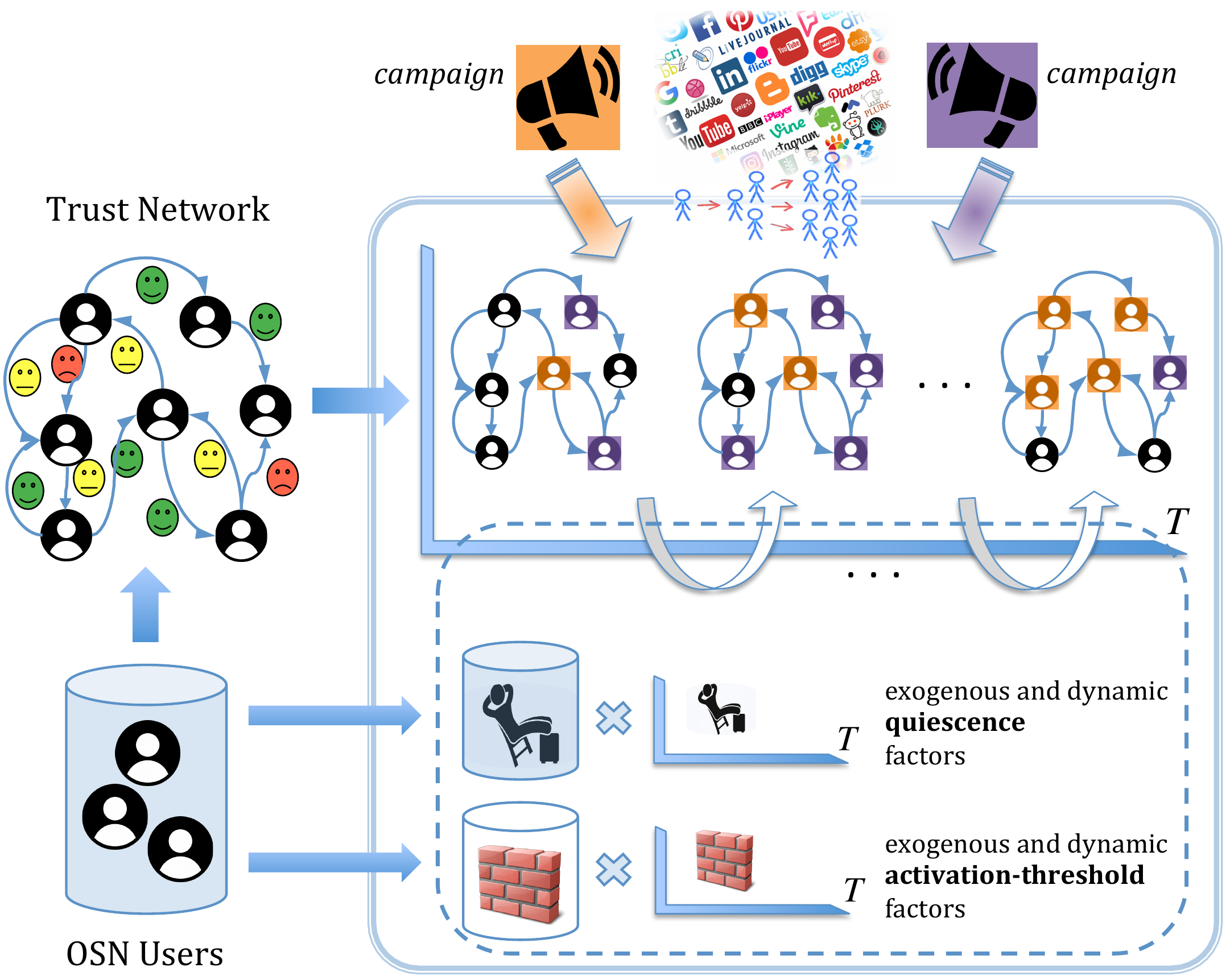}
\caption{Illustration of the information diffusion framework based on our proposed \ffdlt[s].}
\label{fig:illustration}
\end{figure}

 Figure~\ref{fig:illustration} illustrates the conceptual architecture of a framework   based on our proposed models.  
   Given   a population of OSN users,  the framework requires  three   main  inputs: (i)  a \textit{trust network}, which is inferred from the social network of those users to model their trust/distrust relationships;   (ii) user behavioral characteristics that are intrinsic to each user  (i.e., exogenous to an information diffusion scenario) and oriented   to express  two aspects: \textit{activation-threshold}, i.e., the effort needed to activate  a user through cumulative influence from her/his neighbors, and \textit{quiescence}, i.e., the user's hesitation in being  actively committed with the propagation process; and, (iii) one or multiple competing \textit{campaigns}, i.e., information cascades generated from the agent(s) having viral  marketing purposes.   
 Moreover, the information diffusion process has a \textit{time horizon}, and its temporal unfolding  is reflected in the evolution of the information diffusion graph: this also depends on the dynamics of the users' behaviors in response to the influence chains started by the  campaign(s), which admit that users may switch from the adoption of a campaign's item to that of another one.    
   Putting it all together,  our \ffdlt[s] based framework embeds all previously discussed aspects that are required to explain  complex propagation phenomena, i.e., competitive diffusion, non-progressivity,   time-aware activation, delayed propagation,  and  trust/distrust relations.

\subsection{Basic definitions}
We are given a \textit{trust network} represented by a directed
graph $G=\langle V,E,w \rangle$, with set of nodes $V$, set of edges $E$,
and weighting function $w: E \mapsto [-1,1]$ such that, for every  edge $(u,v) \in E$,
$w_{uv} := w(u,v)$ expresses how much  $v$ trusts its in-neighbor $u$. Positive, resp. negative, value of $w_{uv}$ corresponds to a \textit{trust}, resp. \textit{distrust}, relation. 

%
%
 For every $v \in V$, we denote with $N^{in}_{+}(v)$ and $N^{in}_{-}(v)$  the set of neighbors trusted by $v$ (i.e., \textit{friends} of $v$) and the set of neighbors distrusted by $v$ (i.e., \textit{foes} of $v$), respectively.  
Moreover, it holds that  $\sum_{u \in N^{in}_{+}(v)} w_{uv} \leq 1$  and $\sum_{u \in N^{in}_{-}(v)} |w_{uv}| \leq 1$. 

Let $\G = G(g, q, T) = \langle V, E, w, g, q, T \rangle$ be a directed weighted graph representing the LT-based \textit{information diffusion} graph associated with   trust network $G$, where $T$ denotes a \textit{time interval} for the diffusion process,   $g$ and $q$ denote  time-dependent \textit{activation-threshold} and \textit{quiescence} functions.   
These  are   introduced in $\G$ to model the aspects of \textit{time-aware activation} and \textit{delayed propagation}, respectively. 
We   use symbol $\as{}{t}$ to denote  the \textit{set of active nodes} at time $t$, and symbol  $\aqs{}{t}$ to denote  the set of active nodes for which, at $t$,  the quiescence time is not expired yet, i.e., the \textit{quiescent nodes}. 

\paragraph*{\bf Activation-threshold function}
 According to the LT model, every node $v \in V$ is associated with an exogenous activation-threshold, $\theta_v  \in (0,1]$, which corresponds to the a-priori effort needed in terms of cumulative influence to activate the node. 
 We enhance this   concept by defining   an \textit{activation-threshold} function,     $g: V, T \mapsto \mathbb{R}^+$, such that for      $v \in V,t \in T$: 
  $$
  g(v,t) =  \theta_v + \vartheta(\theta_v,t),
  $$
  i.e., the activation of $v$ at time $t$ depends both on the user's pre-assigned  threshold, $\theta_v$,  and on a time-evolving activation term, $\vartheta(\cdot,\cdot)$, which models the dynamic response of a user towards the activation attempt exerted by her/his neighbors.  
   %

To   specify $\vartheta(\cdot,\cdot)$,  we might devise at least two main scenarios for $g(\cdot,\cdot)$: (i)    non-decreasing monotone, and   (ii)   non-monotone function,  
 for any $v \in V$. 
 In the first scenario, $g(\cdot,\cdot)$ would   model the tendency of a user to consolidate her/his belief, according to the  \emph{confirmation-bias} principle~\cite{4ciocchi1}.  
 By contrast,   the second scenario would be useful to capture a user's behavior    to  revise her/his uncertainty to activate over time, thus becoming   more or less inclined to change her/his opinion on an information item. 
   %
   In this work, we focus on the confirmation bias principle, thus choosing the following form for the activation-threshold function, by which the   value  increases by increasing the time a node keeps staying in the same active state:  
\begin{equation}\label{eq:function:activation}
g(v,t) = \theta_v + \vartheta(\theta_v,t) = \theta_v + \delta \times 
\min\left \{\frac{1-\theta_v}{\delta}, t-t_v^{last}\right \}
\end{equation}
%
 where $t_v^{last}$ denotes the last (i.e., most recent) time $v$ 
 was  activated  and $\delta \geq 0$\footnote{We assume the second additive term in Eq. (\ref{eq:function:activation}) is zero if $\delta=0$.}  represents the increment in the value of $g(v,t)$ for consecutive time-steps. 
 Thus, the longer a node has kept its active state for   the same 
 information cascade (\textit{campaign}), the higher its activation  value, and as a consequence,  it will be
harder to make the node  change its state, or even no more possible (i.e., $g(v,t)$ saturates to 1, as the difference $(t - t_v^{last})$ exceeds $(1-\theta_v)/\delta$). 

\paragraph*{\bf Quiescence function}
Each node in $\G$  is also associated with a \textit{quiescence} value, which quantifies the latency in propagation  through that node. 
 We define a \textit{quiescence} function, $q:V,T \mapsto T$, non-decreasing and monotone, such that for every $v \in V, t \in T$, with $v$ activated at time $t$: 
 $$
 q(v,t) = \tau_v + \psi(N^{in}_-(v),t),
 $$
 where $\tau_v \in T$ represents an exogenous term modeling  the  user's hesitation in being fully  committed with  the propagation process, and $\psi(N^{in}_-(v),t)$ 
 provides an additional delay proportional to the amount of $v$'s neighbors that are distrusted and active,  by the time the activation attempt is performed by the $v$'s trusted neighbors: 
 %
\begin{equation}\label{eq:function:quiescence}
q(v,t) = \tau_v + \psi(N^{in}_-(v),t) = \tau_v + \exp\big({\lambda \times  \smashoperator{\sum_{u \in S_{t-1} }} \vert w_-(u,v) \vert }\big)
\end{equation}
where $\lambda \geq 0$ is a   coefficient  modeling the average user sensitivity in the perceived negative influence. Intuitively, this coefficient would weight more the negative influence as the diffusing informative item is more \say{worth of suspicion}.  %

\paragraph*{\bf Rationale for activation and propagation}
Our choice of using, on the one hand,  friends  for the activation   of a user, and on the other hand, foes  to impact on   delayed propagation, represents a key distinction from related work~\cite{realtimecosteffective,Talluri2015479,Weng20171931}. 
In our setting, we tend to reject   as true in general, the principle  
\say{I agree with my friends' idea and disagree with
my foes' idea}, which is also close to the adage \say{the enemy of my enemy is my friend}. 
 Rather,  we believe that 
any user might be provided with a self-determination capability. 
%
 Therefore, in our models, the trusted connections and distrusted connections  play different roles:  only friends can exert a degree of influence, whereas  foes can only contribute to increase the user's hesitation to  commit with  the propagation process. 


\subsection{Non-Competitive Model}
\label{sec:noncomp}
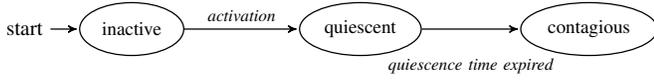
\begin{figure}[t!]
\centering
\scalebox{0.8}{
\begin{tikzpicture}[->,>=stealth',shorten >=1pt,auto,node distance=3.8cm,
                    semithick,transform shape]
  \tikzstyle{every state}=[text=black,draw,ellipse]

  \node[initial,state] (A)                    {\small inactive};
  \node[state]         (B) [right of=A] 	  {\small quiescent};
  \node[state]         (C) [right of=B] 	  {\small contagious};
  
  \path (A) edge              node [above] {\footnotesize \textit{activation}} (B)
        (B) edge 			  node [below=10pt] {\footnotesize \textit{quiescence time expired}} (C) ;
\end{tikzpicture}
}
\caption{Life-cycle of a node in  the \noncomp[s] model.}
\label{fig:ttm:lifecycle}
\end{figure}

We introduce  the first of the three proposed models, which refers to a single-item propagation scenario. Figure~\ref{fig:ttm:lifecycle} shows the   life-cycle  of a node in the diffusion graph under this model. 


\vspace{2mm}
\begin{definition}{\bf \textit{\noncomp} (\noncomp[s]).\ }  
Let $\G =  \langle V, E, w, g, q, T \rangle$ be  the diffusion graph of \noncomp (\noncomp[s]). 
The diffusion process under the \noncomp[s] model unfolds in discrete time steps.  
At time $t=0$, an initial set of nodes $\as{}{0}$ is activated. 
At   time $t \geq 1$, the  following rule applies: for any inactive node   
$v \in V \setminus (\as{}{t-1} \cup \aqs{}{t-1})$,  if
 $\sum_{u \in N^{in}_+(v) \cap S_{t-1}}  w_{uv} \geq g(v,t)$, 
then $v$ will be added to the set of quiescent nodes  $\aqs{}{t}$, 
with   quiescence time equal to $t^* = q(v,t)$. 
Once  the quiescence time is expired, $v$ will be removed  from $\aqs{}{t}$  and  added to the set of active nodes $\as{}{t^*}$.   The process continues until $T$ is expired or  no more 
activation attempts can be performed.
\hfill  \qed 
\end{definition}

 
%

\subsection{Competitive Models}
\label{sec:comp}
Here  we introduce the two competitive \ffdlt[s] models. 
%
Let us first provide our motivation for developing two different competitive 
models:  through the following example, we   
illustrate a  particular situation that may occur when dealing with two 
  campaigns  competitively propagating through a network. 
 Please note that, throughout the rest of this paper, we will consider only two competing campaigns for the sake of simplicity; nevertheless, \textit{our proposed models are generalizable to more than two competing campaigns}.

\begin{figure}[t!]
\begin{tabular}{c@{\hskip 0.22in}c@{\hskip 0.22in}c}
 \begin{tikzpicture}[thick,scale=0.45, every node/.style={scale=0.45},
 	  EdgeStyle/.append style = {->} ]
	  \SetGraphUnit{2}
	  \Vertex[LabelOut=True,Lpos=90,L=$u$]{u}
	  \EA[unit=4,LabelOut=True,Lpos=90,L=$z$](u){z}
	  \SO[LabelOut=True,Lpos=0,L=$x$](z){x}
	  \SOWE[LabelOut=True,Lpos=0,L=$v$](z){v}
	  \AddVertexColor{green}{u,z}
	  \AddVertexColor{red}{x}
	  \Edge[label = 0.3,style={bend left}](u)(v)
	  \Edge[label = 0.5,style={bend right}](z)(v)
	  \Edge[label = 1](x)(z)
	  \node [align=center,anchor=south,below= 0.2 of v] at (v.south west) (8) {$t=0$};
	  \draw (v.west)--++(3:0mm) node [left=0.5] (vparam) {$\theta_v=0.6$};
\end{tikzpicture}

&

  \begin{tikzpicture}[thick,scale=0.45, every node/.style={scale=0.45},
  	  EdgeStyle/.append style = {->} ]
 	  \SetGraphUnit{2}
 	  \Vertex[LabelOut=True,Lpos=90,L=$u$]{u}
 	  \EA[unit=4,LabelOut=True,Lpos=90,L=$z$](u){z}
 	  \SO[LabelOut=True,Lpos=0,L=$x$](z){x}
 	  \SOWE[LabelOut=True,Lpos=0,L=$v$](z){v}
	  \AddVertexColor{green}{u,v}
	  \AddVertexColor{red}{x,z}
	  \Edge[label = 0.3,style={bend left}](u)(v)
	  \Edge[label = 0.5,style={bend right}](z)(v)
	  \Edge[label = 1](x)(z)

	  \draw (v.west)--++(3:0mm) node [left=0.5] (vparam) {$\theta_v=0.6$}; 
	  \node [align=center,anchor=south,below= 0.2 of v] at (v.south west) (8) {$t=1$};
\end{tikzpicture}

&

 \begin{tikzpicture}[thick,scale=0.45, every node/.style={scale=0.45}, VertexStyle/.append style ={pattern=north east lines, inner sep=0pt},
 	  EdgeStyle/.append style = {->} ]
	  \SetGraphUnit{2}
	  \Vertex[LabelOut=True,Lpos=90,L=$u$]{u}
	  \EA[unit=4,LabelOut=True,Lpos=90,L=$z$](u){z}
	  \SO[LabelOut=True,Lpos=0,L=$x$](z){x}
	  \SOWE[LabelOut=True,Lpos=180,L=$v$](z){v}
	  \AddVertexColor{green}{u}
	  \AddVertexColor{red}{x,z}
	  \Edge[label = 0.3,style={bend left}](u)(v)
	  \Edge[label = 0.5,style={bend right}](z)(v)
	  \Edge[label = 1](x)(z)
	  \node [align=center,anchor=south,below= 0.2 of v] at (v.south west) (8) {$t=2$};
	  \node [align=center,anchor=east,above= 1.3 of v] at (v.south) (quest) {\textit{Keep green-active or  	deactivate?}};
	  \draw[dashed] (v)--(quest);
  	  \draw (v.west)--++(3:0mm) node [left=0.5] (vparam) {$\theta_v=0.6$};
\end{tikzpicture}
 
\end{tabular}
\caption{Uncertainty in a two-campaign activation sequence.} 
\label{fig:activation_sequence}
\end{figure}
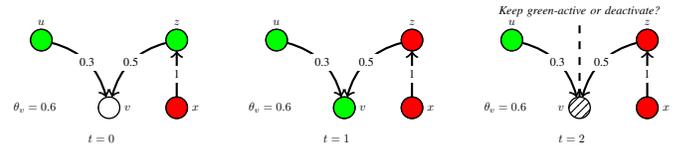

\begin{example}\label{exe:activation_sequence}
 Figure~\ref{fig:activation_sequence} shows an example activation sequence in a competitive scenario between two information cascades, distinguished by colors red and green.   
At time $t=0$, nodes $u$ and $z$ are green-active, and their joint influence causes green-activation of   node $v$ as well (since $0.3+0.5 \geq 0.6$). 
 At time $t=1$, as fully influenced by node $x$, node $z$ has switched its activation in favor of the red campaign. After this switch, at time $t=2$, it happens that $v$'s activation  status is no more consistent with the (joint or individual) influenced exerted by   $u$ and $z$. In particular, two mutually exclusive events might in principle happen at $t=2$: either $v$ is deactivated or $v$ maintains its green-activation state.
 \hfill $\blacksquare$
\end{example}

The uncertainty situation depicted in the above example  prompted us to  
the definition of two models, namely \emph{semi-progressive} and   \emph{non-progressive}  \ffdlt[s]: the former corresponds to the case of $v$ keeping its current (i.e., green)  activation state, whereas the latter corresponds to $v$ returning to the inactive state.  
  %
 Clearly, the two models' semantics are different from each other: the semi-progressive model assumes that a user, once activated, cannot step aside, unlike the non-progressive one, which instead requires a user to have always the support of her/his in-neighbors to keep activation. 

 Given two information cascades, or \textit{campaigns}  $C', C''$, for every time step $t \in T$  we will use symbols  $\as{+}{t}$ and $\as{-}{t}$ to denote   the sets of active nodes, such that that $\as{+}{t} \cap \as{-}{t} = \emptyset$, and analogously symbols  $\aqs{+}{t}$ and $\aqs{-}{t}$ as the sets of quiescent nodes, for   $C'$ and $C''$,  
respectively. Also,   $\as{}{t} = \as{+}{t} \cup \as{-}{t}$ and $\aqs{}{t} = \aqs{+}{t} \cup
\aqs{-}{t}$.

It should also be noted that, while sharing  the  time interval ($T$) of diffusion,  $C'$ and $C''$ are not constrained to start at the same time $t_0$. Nevertheless, for the sake of simplicity, we hereinafter assume that $t_0=t_0'=t_0''$ (with $t_0 \in T$), unless otherwise specified (cf. Sect.~\ref{sec:results}).


\vspace{2mm}
\begin{definition}{\bf \textit{\semiprog} (\semiprog[s]).\ }
Let $\G =  \langle V, E, w, g,$  $q, T \rangle$ be  the diffusion graph of \semiprog (\semiprog[s]), and  $C', C''$ be  two   campaigns  on $\G$. 
The diffusion process under the \semiprog[s] model unfolds in discrete time steps. At time $t=0$, two initial sets of nodes, $\as{+}{0}$ and   $\as{-}{0}$, are activated for each campaign. At every time step $t \geq 1$, the following rules apply:

{\bf R1.\ } 
For any inactive node $v \in V \setminus (\as{}{t-1} \cup
\aqs{}{t-1})$, \\ if 
$\sum_{N^{in}_+(v) \cap \as{+}{t-1}} w_{uv} \geq g(v,t)$, 
then $v$ will be added to $\aqs{+}{t}$; analogous rule holds for $C''$.   

{\bf R2.\ } 
Given  a  node active for $C''$,  $v \in \as{-}{t-1}$, if 
 $\sum_{N^{in}_+(v) \cap \as{+}{t-1}} w_{uv}   \geq   g(v,t)$ and $\sum_{N^{in}_+(v) \cap \as{+}{t-1}} w_{uv}   >   \sum_{N^{in}_+(v) \cap \as{-}{t-1}} w_{uv}$,  
then $v$ will be removed from $\as{-}{t}$ and added to $\as{+}{t}$; analogous rule holds for   any  node active for the first campaign.    

{\bf R3.\ } 
Every active node for which none of the above conditions
is matched will keep its current state.

{\bf R4.\ } 
When a node $v$ is activated for the first time, it will stay 
in $\aqs{+}{t}$ or $\aqs{-}{t}$ until the quiescence  
time is   expired. 

{\bf R5.\ } 
For every node that can be  simultaneously activated
by both campaigns,  a \textit{tie-breaking} rule will apply, in order to decide which campaign actually determines the node's activation.
\hfill \qed 
\end{definition}

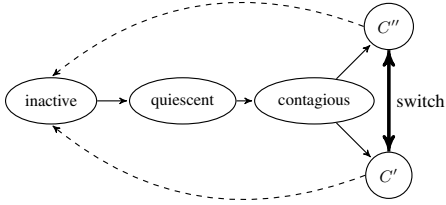
\begin{figure}[t!]
\centering
%
%
 \scalebox{0.7}{
\begin{tikzpicture}[->,>=stealth',shorten >=1pt,auto,node distance=2.5cm,
                    semithick]
  \tikzstyle{every state}=[text=black,draw,ellipse]

  \node[state]         (A)              	  {\small inactive};
  \node[state]         (B) [right of=A] 	  {\small quiescent};
  \node[state]         (C) [right of=B] 	  {\small contagious};
  \node[state]		   (D) [above right of=C,node distance=2cm]  {\small $C''$};
  \node[state]		   (E) [below right of=C, node distance=2cm] {\small $C'$};

  \path (A) edge              	node {} (B)
        (B) edge 			 	node {} (C) 
        (C) edge			  		node {} (D) 
        (C) edge              	node {} (E)
        (D) edge [line width=1.7]  node {} (E)
        (E) edge [line width=1.7, right]  node {switch} (D)
        (E.west) edge[bend left,dashed]   node {} (A.south)
        (D.west) edge[bend right,dashed]  node {} (A.north);
\end{tikzpicture} 
}
\caption{Life-cycle of a node in competitive models. Dashed lines are valid for \nonprog[s] only.}
\label{fig:compm:lifecycle}
\end{figure}

As shown  in Fig.~\ref{fig:compm:lifecycle},      once a node becomes active, 
it cannot turn back to the inactive state, but it can only change the activation campaign. 
Moreover, switch transitions occur instantly. 

\vspace{2mm}
\begin{definition}{\bf \textit{\nonprog} (\nonprog[s]).\ } 
Let $\G =  \langle V, E, w, g,$  $q, T \rangle$ be  the diffusion graph of \nonprog (\nonprog[s]), and  $C', C''$ be  two   campaigns  on $\G$.   
The diffusion process in    \nonprog[s]    evolves according to the same rules as in \semiprog[s] plus  the following rule concerning the deactivation process of an active node:

{\bf R6.\ } 
For any active node $v$ at time $t-1$, if 
 $\sum_{N^{in}_+(v) \cap \as{+}{t-1}} w_{uv}   <   \theta_v$ and $\sum_{N^{in}_+(v) \cap \as{-}{t-1}} w_{uv}  <  \theta_v$, 
then $v$ will turn back to the inactive state at time $t$.
\hfill \qed 
\end{definition}

 It should be noted that a node's deactivation rule depends on      $\theta_v$ only (rather than on the whole function $g(v,t)$); otherwise, every node activated at a given time  could deactivate itself  in the next time step, due to the increase in its activation threshold.     
 %
    In Fig.~\ref{fig:compm:lifecycle}, note that, unlike in    \semiprog[s],  transitions to inactive state are allowed.

\section{Evaluation methodology}
\label{sec:eval}

\subsubsection*{\bf Data}
We used four real-world, publicly available networks, namely: 
\textit{Epinions}~\cite{Leskovec20101361}, 
\textit{Slashdot}~\cite{Leskovec20101361}, 
\textit{Wiki-Conflict}~\cite{konect:brandes09}, 
and \textit{Wiki-Vote}~\cite{konect:leskovec207}.  
 The first two are \say{who-trust-whom} networks,  
Wiki-Conflict refers to Wikipedia users involved in an \say{edit-war} (i.e.,  edges represent either positive  or negative conflicts in editing a wikipage), and  
Wiki-Vote models   relations between  Wikipedia users that voted for/against each other in admin elections. 
Table~\ref{tab:graph:properties} summarizes main characteristics of the networks. Note that 
   to favor meaningful competition  of  campaigns based on selected pairs of strategies, we limited the diffusion context to the largest strongly connected component in each evaluation network, except for Wiki-Conflict.


All networks are originally directed and signed; in addition, the two Wikipe\-dia-based networks also have timestamped edges. 
In order to derive the weighted graphs of influence probabilities,  we defined the following method: 
for every $(u,v) \in E$, the edge weight $w_{uv}$ was sampled from a binomial distribution $\mathcal{B}(|N^{in}_+(v)|,p)$  if $u \in N^{in}_+(v)$ (i.e., $v$ trusts $u$), otherwise  $w_{uv} \sim -\mathcal{B}(|N^{in}_-(v)|,p)$, where the probability of success $p$ is equal to the fraction of trust edges in the network.   
  %
 %
    We performed $1,000$ samplings of edge weights, for each of the four networks. Therefore, all presented results will correspond to  averages of $1,000$ simulation runs.

\subsubsection*{\bf Seed selection strategies}
We defined four   seed   selection strategies, each of which   mimics  a different, realistic scenario of influence propagation. 
 
\paragraph*{Exogenous and malicious sources of information}
This method, hereinafter referred to as  \textsf{M-Sources}, aims at simulating the presence of multiple sources of malicious information within the network. Here, an exogenous source is meant as a node without incoming links, e.g., a user that is just interested in spreading her/his opinion: such a node is also regarded as malicious if a high fraction of outgoing influence exerted by the node is distrusted by out-neighbors. Formally, given a budget $k$, the method  selects the top-$k$ users in a ranking solution determined as      
$r(v) = (\bar{W}^- / (\bar{W}^- + \bar{W}^+)) \log(|N^{out}(v)|)$, for every $v$ such that $N^{in}(v)=\emptyset$, where $\bar{W}^+, \bar{W}^-$ are shortcut symbols to denote the sum of trust (resp. distrust) weights, respectively, outgoing from $v$. 
%

\paragraph*{Exogenous and influential trusted sources of information}
Analogously to the previous method, this one, dubbed \textsf{I-Sources},  searches for the \say{best}  influential trusted  sources. The ranking function is as  $r(v) = (\bar{W}^+ / (\bar{W}^- + \bar{W}^+)) \log(|N^{out}(v)|)$. Note that this  still  takes into account the negative weights, because even a highly trusted  user might be distrusted by some other users  (e.g., \say{haters}).

\begin{table}[t!] 
\centering 
\caption{Summary of evaluation network data.}
\label{tab:graph:properties}
\scalebox{0.8}{
\begin{tabular}{|l|c|c|c|c|}
\hline
 & \textit{Epinions} & \textit{Slashdot} & \textit{Wiki-Conflict} & \textit{Wiki-Vote} \\
\hline \hline 
{\#nodes} & 131\,828  & 77\,350   & 116\,836  & 7\,118	\\
{\#edges} & 841\,372 	& 516\,575 	& 2\,027\,871 & 103\,675 \\
 \% distrusted/negative-edges   & 14.7\% & 23.3\% & 61.9\% & 21.6\% \\ 
avg. degree &  6.38  & 6.67  & 17.36 &  6.68 \\
diameter & 14 & 11 & 10 & 7 \\
clust. coeff. & 0.093 &	0.026 & 0.015 & 0.128 \\
\textit{strong LCC} {\#nodes} &  36\,490 & 23\,217 & -- & 1\,178  \\ 
\textit{strong LCC} {\#edges} &  602\,722 & 243\,600 & -- & 31\,572	\\
\hline
\end{tabular}
} 
\end{table}

\paragraph*{Stress triads}
This strategy is based on the notion of \textit{structural balance} in triads~\cite{Leskovec20101361}. 
 Suppose   node $v$ has two incoming connections, the one   from node $z$ with negative weight, 
 and the other   from $u$ with  positive weight; moreover, there is   a trust link from $z$ to $u$. 
 We say that $z$ is a \textit{stress-node}  since, despite the distrusted link to $v$, it could   indirectly  influence  $v$ through the trusted connection with $u$. 
 Our proposed   \textsf{Stress-Triads} strategy searches for all triads containing stress-nodes and selects as seeds the first $k$ stress-nodes with the highest number of triads they participate  to. 

\paragraph*{Newcomers}
 We call a node   
$v \in V$ as a \textit{newcomer} if  
 all of its incoming edges   are timestamped as less recent than  its oldest outgoing edge.   
  The \textit{start-time} of $v$ is the oldest timestamped associated with its incoming edges. 
 We divide the set of newcomers into two groups obtained by equal-frequency binning on the temporal range specific of a network.  Upon this, we distinguish  between two strategies, dubbed \textsf{Least-New} and  \textsf{Most-New}, which correspond to the selection of $k$ newcomers having highest out-degree   among those with  the oldest start-time and with the newest start-time, respectively. 
 %
 Both strategies were applied to Wiki-Vote and   Wiki-Conflict, due to the availability of timestamped edges.

 \begin{figure}[t!]
 \begin{tabular}{cc}
 \includegraphics[height=3.2cm, width=3.8cm]{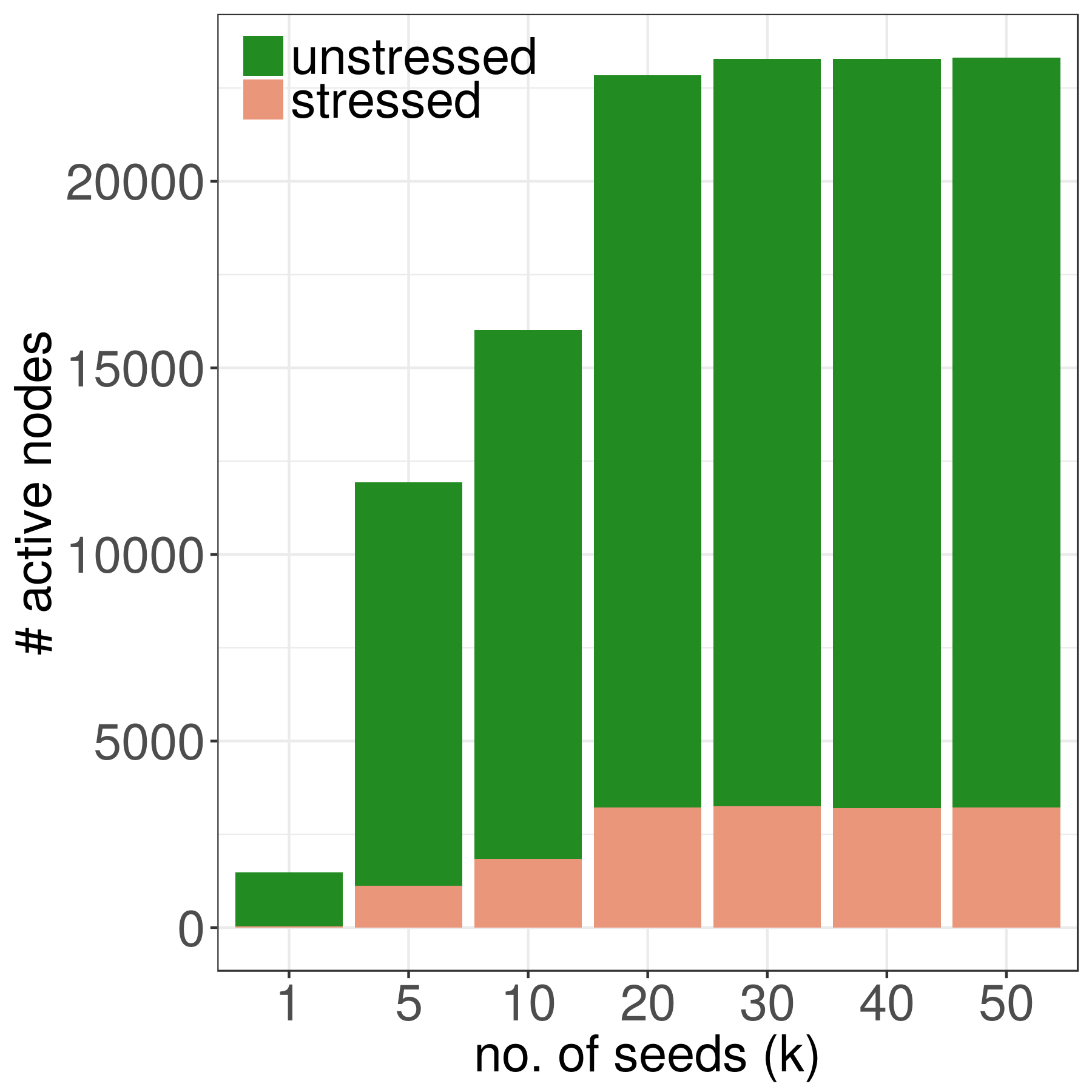} & 
  \includegraphics[height=3.2cm, width=3.8cm]{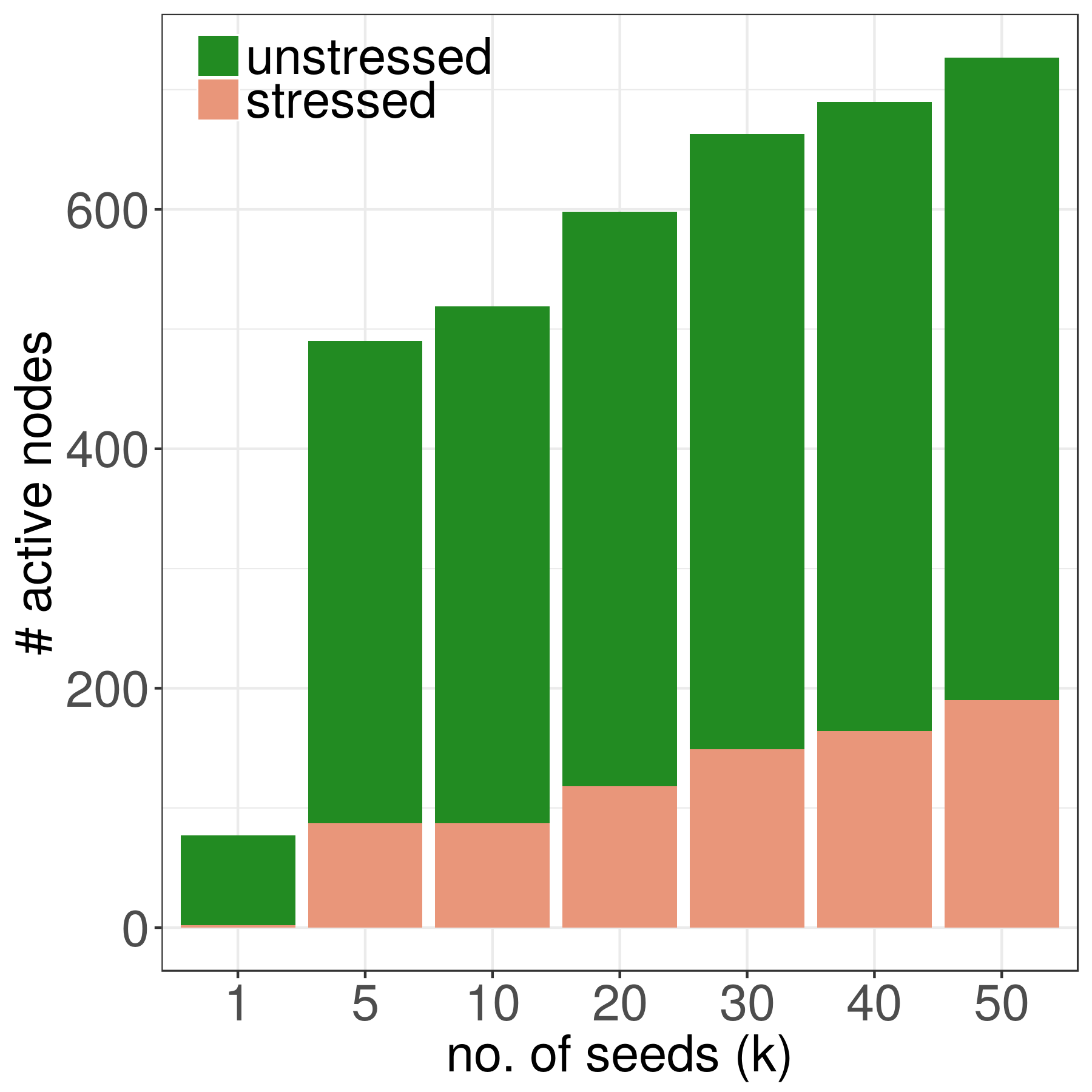} 
 \\
 (a) Epinions, \textsf{I-Sources} 
 & (b) Wiki-Vote, \textsf{Stress-Triads} \\
 \end{tabular}
 \caption{Spread of \noncomp[s] by varying seed set size ($k$) and selection strategy.}
 \label{fig:nc-activation}
 \end{figure}

\subsubsection*{\bf Settings of the model parameters}
For every user $v$, the exogenous activation-threshold  $\theta_v$  and   quiescence time  $\tau_v$   were chosen uniformly at random within [0,1] and [0,5]. 
Moreover, 
   $\lambda$  and   $\delta$  were  varied  between 0 and 5, and 
      between 0 and 0.5, respectively. 
    
    


\section{Results}
\label{sec:results}
  
\subsection{Evaluation of \noncomp[s]}
 \label{sec:eval-noncompetitive}

\paragraph*{\bf  Spread, stressed users}
 We analyzed the number of final activated users (i.e., \textit{spread}) by varying the size ($k$) of seed set, for every seed selection strategy. 
 We initially assumed constant activation thresholds (i.e., $\vartheta(\cdot, \cdot)=0$) and constant quiescence times (i.e., $\psi(\cdot, \cdot)=0$). Moreover, we   distinguished between \say{\textit{stressed}} and \say{\textit{unstressed}} users, being the former regarded as active users having at least one distrusted active in-neighbor. 
 As shown in Fig.~\ref{fig:nc-activation} for some representative cases,   
    we found the activation of 
    the stressed users were activated less than the unstressed users, although with similar trend as $k$ increases.  
  \textsf{I-Sources} along with \textsf{Stress-Triads} revealed higher  spread capability,  in all networks (with the exception of Wiki-Vote).  
   \textsf{Least-New} prevailed  on \textsf{Most-New}  for lower $k$.  
     \textsf{M-Sources} led to a much lower  spread than the other strategies. 
 \paragraph*{\bf Activation loss} 
We further investigated the \textit{activation loss}, i.e., the percentage decrease of activated users due to the enabling of  the time-varying quiescence factor (i.e., $\lambda>0$ in Eq.~\ref{eq:function:quiescence}).   
 %
  %
  By setting a relatively large  $\lambda$ (set to 5) and $k$ (set to 50), we found high percentage of activation loss for the initial time steps, especially for  \textsf{Stress-Triads}, which might be explained since the initial influenced  users tend to be subjected to a certain amount of distrusted influence.  
  As  the time horizon approaches, the activation loss tends to significantly decrease, down to nearly zero in most cases, with  few   exceptions including the use of \textsf{I-Sources} in Slashdot and Epinions, and \textsf{Stress-Triads} and \textsf{M-Sources} in Wiki-Vote.

\begin{table}[t!] 
\centering 
\caption{Statistics about selected pairs of strategies for two campaigns: the seed set $S_0^{(1)}$ (resp. $S_0^{(2)}$) computed for the first-started or ``bad''  (resp. second-started or ``good'') campaign $SS_1$ (resp. $SS_2$), the   spread $|\Phi(S_0^{(1)})| (resp. |\Phi(S_0^{(2)})|$),  
 the fraction of spread of the bad campaign shared with the good  campaign (\textit{shared} column), the percentage of shared users that were activated first by the bad campaign (\textit{$SS_1$ first} column), the average time of activation of the shared users, and   the average time of activation of the shared users by the bad campaign before the good campaign, and vice versa.  
 Abbreviations \textsf{IS}, \textsf{MS}, \textsf{ST}, \textsf{LN}, and \textsf{MN} stand for \textsf{I-Sources}, \textsf{M-Sources}, \textsf{Stress-Triads}, \textsf{Least-New}, and \textsf{Most-New}. 
}
\label{tab:competitive-info}
\scalebox{0.78}{
\begin{tabular}{|l|c|c|c|c||c|c|c|c|c|c|}
\hline
network & $SS_1$ & $SS_2$ &\!\!$|\Phi(S_0^{(1)})|$\!\!&\!\!$|\Phi(S_0^{(2)})|$\!\!& 
	shared	
        & $SS_1$  & \multicolumn{3}{|c|}{avg. activation time} 
         \\
         \cline{8-10} 
        & & & & & & first &   any & $SS_1$  & $SS_2$   \\ 
         & & & & & & &    &first & first  \\
\hline
\hline

\multirow{2}*{\textit{Epinions}}\!\!& \textsf{ST} 	& \textsf{IS}	& 10595 & 23321 
& 0.99 &   28\%	 & 6.03 & 0.67 	& 5.27   \\
								& \textsf{MS}		& \textsf{IS}	& 59	& 23321	
								& 0.01 &   100\% & 4.0	& 3.0	&	0.0  \\
\hline
\multirow{2}*{\textit{Slashdot}}\!\!& \textsf{ST} 	& \textsf{IS}	& 3263 & 18671 
 & 0.98 &   40\%	& 6.56 	&  2.54	& 7.63   \\
								& \textsf{MS}     	& \textsf{IS}	& 58   & 18671 
								 & 0.05 &	  100\%	& 4.0   &  5.0  & 0.0 	 \\
\hline
\multirow{2}*{\textit{Wiki-}}\!\!& \textsf{ST} 	& \textsf{IS}   	& 344  & 5968 
&  0.84 & 	95\% 
             & 2.43 	&  1.43	& 4.93   \\
			 	& \textsf{MS} 		& \textsf{IS}   	& 203  & 5968 
								&  0.75 & 	98\%	& 3.64 	&  0	& 9.5   \\
 				\textit{Conflict}\!\!& \textsf{LN}  	& \textsf{MN} 	& 216  & 424  
 								&  0.7  & 	100\%	& 3.77 	&  0	& 0   \\  		
\hline                                
\multirow{2}*{\textit{Wiki-}}\!\!& \textsf{ST}	& \textsf{IS}	& 727	 & 394 
 & 0.45	&  78\%    & 4.32  & 0.54  & 5.87 \\
								& \textsf{MS}		& \textsf{IS}	& 172  	 & 394 
								& 0.41	&  79\%    & 4.04  & 0.05  & 3.93 \\
                     \textit{Vote}           & \textsf{LN}		& \textsf{MN} 	& 165	 & 159 
                                  & 0.13  &  63\%	   & 2.72  & 0.0   & 4.37 \\

\hline
\end{tabular}
}
\end{table}

 \subsection{Evaluation of competitive models}
 \label{sec:eval-competitive}
To  analyze  the behavior of  competitive models,   
  we simulated   a scenario of \textit{limitation of misinformation spread}, i.e., we assumed that      the \say{bad} campaign has started diffusing, and  the \say{good} campaign   is carried out in reaction to   the first one. 
 To this end,     
   preliminarily to this evaluation,  
  we   investigated about proper combinations of  seed selection strategies.  
   %
     Table~\ref{tab:competitive-info} provides  statistics about  selected pairs of   strategies, for two campaigns carried out independently to each other,  with $k=50$.   %
  %
   We observe that 
   using \textsf{Stress-Triads} and \textsf{I-Sources} for the bad  and good campaigns, respectively, is particularly  significant, with sharing   close to 100\% in Epinions and Slashdot and above 80\% in Wiki-Conflict.   

 In the following, we present results aimed to understand  the effect of the confirmation bias factor 
  on the users' campaign-changes/deactivations, under the case of \say{real-time correction} or \say{delayed correction} by the good campaign against the bad one (cf. Introduction).   
     We used  fixed-probability as tie-breaking rule, with probability  1 for the bad campaign, and we set the time horizon to the end-time of the (non-competitive) diffusion  of the bad campaign. 

\begin{figure}[t!]
 \begin{tabular}{cc}
 \hspace{-4mm}
 \includegraphics[height=3cm,width=0.24\textwidth]{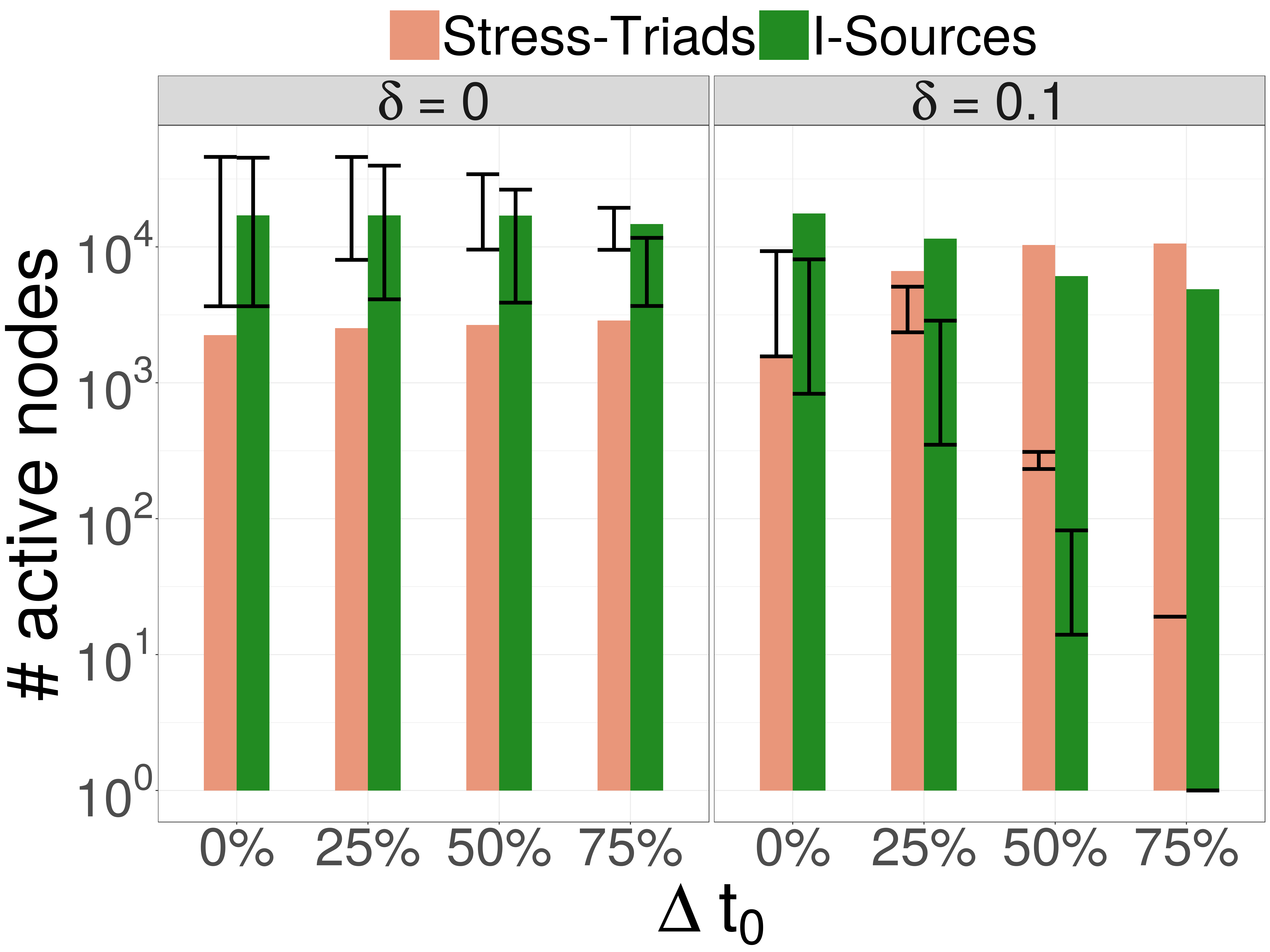} &  \hspace{-2mm}
 \includegraphics[height=3cm,width=0.24\textwidth]{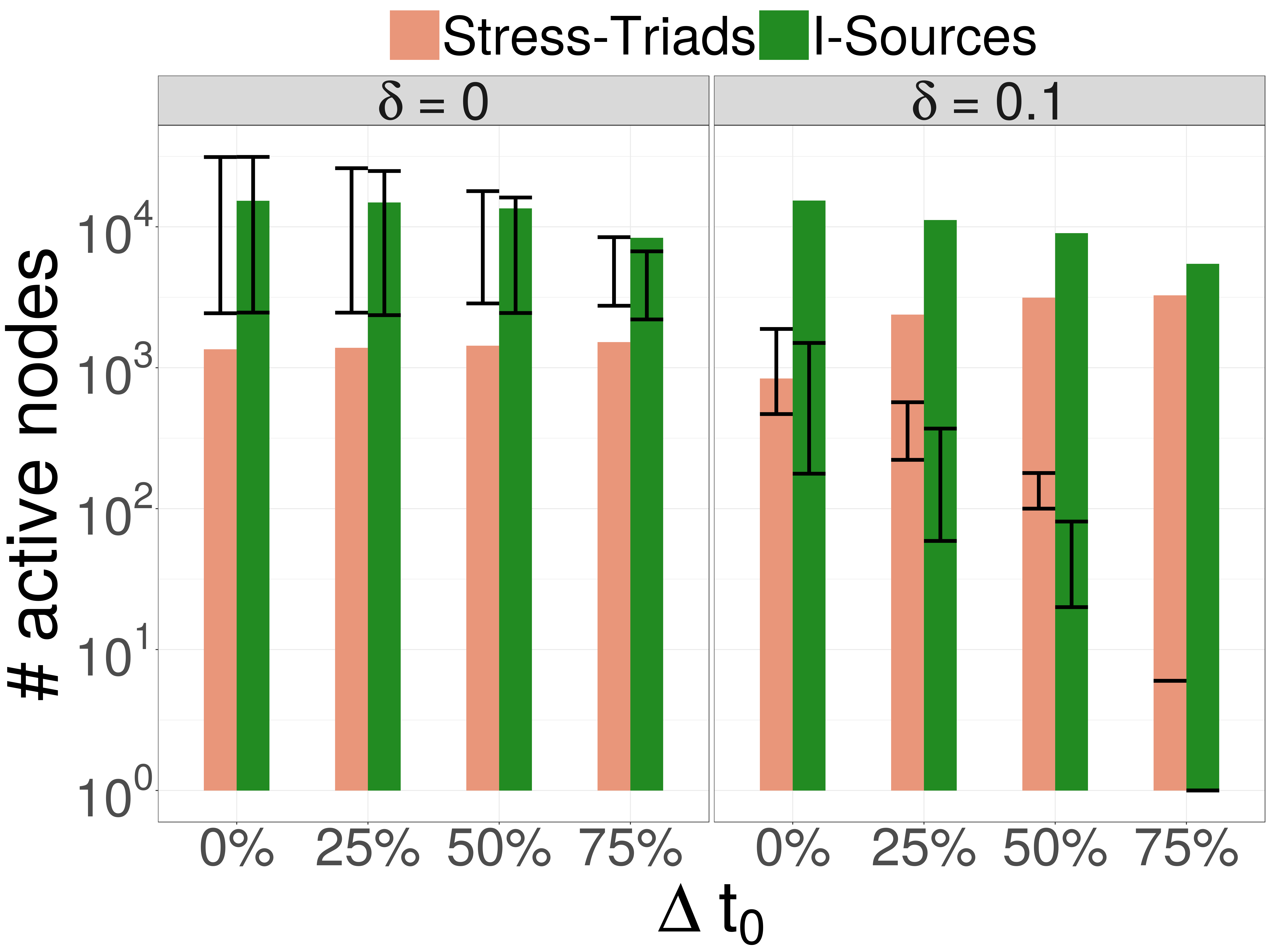} \\
 \hspace{-4mm} (a) Epinions & (b) Slashdot    \\ 
 \hspace{-4mm} \includegraphics[height=3cm,width=0.24\textwidth]{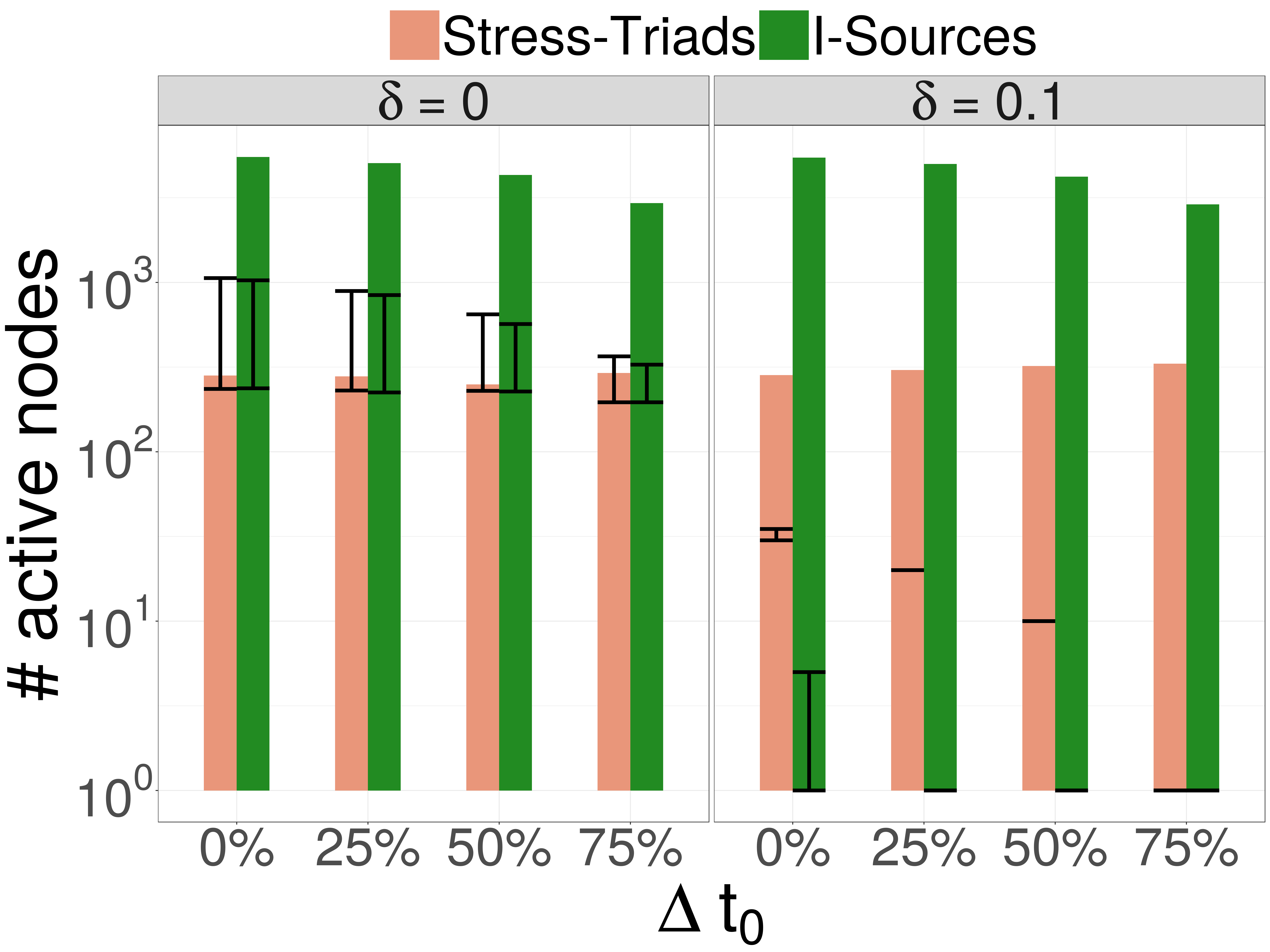} & \hspace{-2mm} \includegraphics[height=3cm,width=0.24\textwidth]{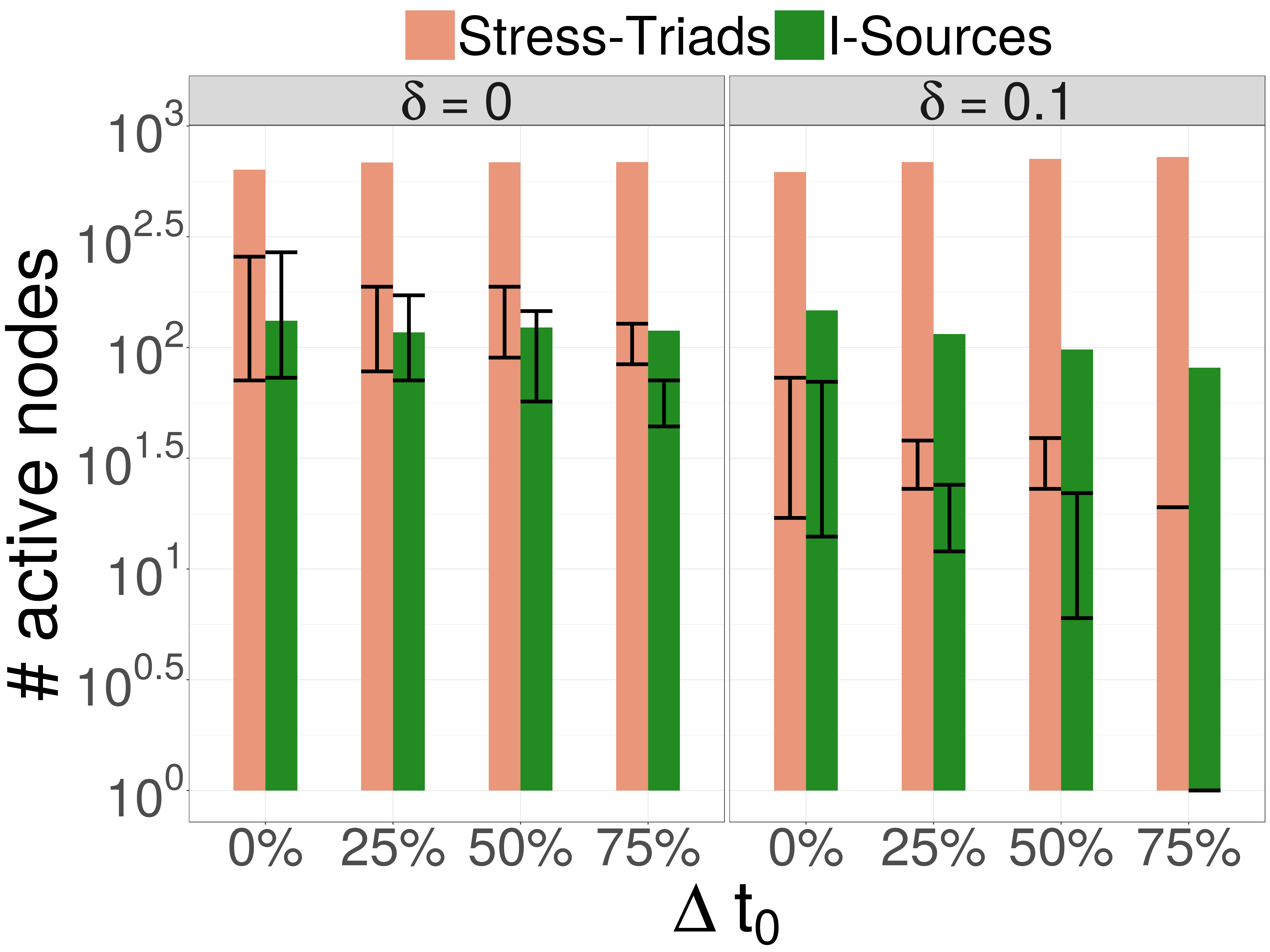} \\
 \hspace{-4mm}  (c) Wiki-Conflict  & (d) Wiki-Vote     \\
 \end{tabular}
 \caption{\semiprog[s]: Spread, number of switched users (lower whiskers), and number of switches (lower whiskers), in log-scale, by varying start-delay ($\Delta\,t_0$) of the ``good''  campaign (second bars), for $\delta=\{0,0.1\}$ and $k=50$. 
 }
 \label{fig:sp-spreads}
 \end{figure}

\paragraph*{\bf Evaluation of \semiprog[s]}
 Figure~\ref{fig:sp-spreads}  shows results on the campaign spreads, 
     the number of users activated for one campaign that \textit{switched} to the other campaign, and the total number of switches, by varying $\delta$ and start-delays  $\Delta\,t_0$  of the good campaign  (up to 75\% of the time horizon).     

 Focusing on the campaign-switches,  
 for $\delta=0$, the number of switched users    follows a nearly constant trend in most networks,  as the start-delay increases, 
     while the total number of switches is  subjected to a more evident decreasing trend. Also,  we observe  a higher number of   (unique and total) switches from  the bad campaign to the good one, than vice versa. 
      Setting $\delta=0.1$ leads to a   general decrease in the switch measurements w.r.t. the
corresponding   case for $\delta=0$.   

 


\paragraph*{\bf Evaluation of \nonprog[s]} 
 The spread trends observed under \nonprog[s] are similar to those corresponding to  \semiprog[s] but, more importantly,    the occurrence of deactivation events, which are admitted by \nonprog[s],  appeared to favor the good campaign strategy.   
  In particular, with combinations      \textsf{Stress-Triads} or \textsf{M-Sources} vs. \textsf{I-Sources}, the number of user-unique and total deactivations tend to increase (resp. decrease)  for the bad (resp. good) campaign as $\Delta\,t_0$ increases; also, for $\delta>0$, the spread of the good campaign would remain higher than the spread of the bad one, due to a much larger number of deactivations from the bad campaign, up to one order of magnitude in Epinions and Slashdot, or even two orders  of magnitude in Wiki networks.   
   %
%

   \subsection{Lessons Learned}
The results of our evaluation revealed that the average user's sensitivity in the  negative influence perceived from distrusted neighbors (which is controlled by  $\lambda$)   
 makes the seed identification process more  aware of the negative influence spread, thus considering the quiescence-biased contingencies by which 
 a non-negligible fraction of users cannot be activated before the time limit.   
%
 
 The  confirmation-bias effect underlying $\delta$  may lead  the  \say{strong\-er} campaign (i.e., the one able to activate most users at the early   steps of its diffusion)   to   increase its spread capability.

 When using the semi-progressive competitive model (\semiprog[s]),  the combined effect of increased $\delta$ with an increase in the delay of the beginning of the second-started (good) campaign may reduce its  capability of \say{saving} users from the influence of the bad campaign; therefore, to limit misinformation spread, the good campaign should concentrate its (activation) efforts in the early stage of its diffusion.  
 Nonetheless, the non-progressive competitive model (\nonprog[s]) appears to be more robust   to the increase of $\delta$, in favor of the good campaign. Yet, \nonprog[s] tends to favor deactivation events (for users previously activated by  the weaker campaign) over switched events. 
 Overall, this would suggest that the misinformation limitation problem could be more easily  addressed by allowing  users to ``reset'' their opinion when biased by the bad campaign, before eventually adopting the good campaign's choice.

\section{Conclusions} 
We proposed a novel class of trust-aware, dynamic LT-based models for non-competitive and competitive influence propagation. Evaluation on real-world, publicly available networks  included simulations of scenarios of misinformation spread limitation, based on realistic strategies of selection of the initial influential users.   
 Our models pave the way for the development of sophisticated methods to solve  misinformation spread limitation and related optimization problems.

Further information on this research work can be found at {\em \url{http://people.dimes.unical.it/andreatagarelli/ffdlt/}}.
 
 \bibliographystyle{plain}
  \bibliography{backend}

\end{document}